\documentclass[iop]{emulateapj}

\usepackage{apjfonts}
\usepackage[T1]{fontenc}
\usepackage{color}
\usepackage{amsmath,amstext}
\usepackage{hyperref}
\usepackage{natbib}
\usepackage{graphicx}
\usepackage{gensymb}
\graphicspath{{./images/}}
\begin{document}

\title{Intrinsic properties of the engine and jet
that powered the short gamma-ray burst associated with GW170817}
\author{Davide Lazzati$^1$, Riccardo Ciolfi$^{2,3}$, Rosalba Perna$^{4,5}$}

\affil{$^1$ Department of Physics, Oregon State University, 301
  Weniger Hall, Corvallis, OR 97331, USA} 
\affil{$^2$ INAF, Osservatorio Astronomico di Padova, Vicolo dell'Osservatorio 5, I-35122 Padova, Italy}
\affil{$^3$ INFN, Sezione di Padova, Via Francesco Marzolo 8, I-35131 Padova, Italy}
\affil{$^4$ Department of Physics and Astronomy, Stony Brook
  University, Stony Brook, NY 11794-3800, USA}
\affil{$^5$ Center for Computational Astrophysics, Flatiron Institute, New York, NY 10010, USA} 

\begin{abstract}
GRB\,170817A was a subluminous short gamma-ray burst detected about
$1.74$\,s after the gravitational wave signal GW170817 from a binary
neutron star (BNS) merger. It is now understood as an off-axis event
powered by the cocoon of a relativistic jet pointing 15 to 30 degrees
away from the direction of observation. The cocoon was energized by the
interaction of the incipient jet with the non-relativistic baryon wind
from the merger remnant, resulting in a structured outflow with a narrow
core and broad wings. In this paper, we couple the observational
constraints on the structured outflow with a model for the jet-wind
interaction to constrain the \textit{intrinsic} properties with which
the jet was launched by the central engine, including its time delay
from the merger event. Using wind prescriptions inspired by magnetized
BNS merger simulations, we find that the jet was launched within about
$0.4$\,s from the merger, implying that the $1.74$\,s observed delay was
dominated by the fireball propagation up to the photospheric radius. We
also constrain, for the first time for any gamma-ray burst, the jet
opening angle at injection and set a lower limit to its asymptotic
Lorentz factor. These findings suggest an initially Poynting-flux
dominated jet, launched via electromagnetic processes. If the jet was
powered by an accreting black hole, they also provide a significant
constraint on the survival time of the metastable neutron star remnant.
\end{abstract}

\section{Introduction}
\label{intro}

The discovery of the gravitational wave source GW170817
\citep{LVC2017-BNS} marked the first detection of gravitational waves
(GWs) from a binary neutron star (BNS) merger.  The observation of the
same source in the electromagnetic spectrum, from the almost
simultaneous $\gamma$-rays
\citep{LVC2017-GRB,Goldstein2017,Savchenko2017} to the later X-ray and
UV, optical, IR, and radio signals \citep{LVC2017-MMA}, allowed, among
other astrophysical implications, to firmly establish the connection
between short gamma-ray bursts (SGRBs) and BNS mergers (e.g.,
\citealt{LVC2017-GRB,Goldstein2017,Savchenko2017,Troja2017,Hallinan2017,
Kasliwal2017,Lazzati2018,Ghirlanda2019,Mooley2018}).
    
The early UV, optical, and IR radiation, detected within about a day
from the GW/$\gamma$-ray detection, were shown to be consistent, both
spectrally and temporally, with the expectations of a kilonova (e.g.,
\citealt{Arcavi2017,Soares-Santos2017,Pian2017}), i.e.~a transient
powered by the radioactive decay of heavy r-process elements synthesized
within the matter ejected during and after merger. The later X-ray
\citep{Troja2017} and radio emission \citep{Hallinan2017}, first
detected $\gtrsim10$ days after the trigger, followed a single power-law
spectrum over more than eight orders of magnitude in energy
\citep{Lyman2018}. This suggested an origin in a blastwave, and the
spectral-temporal characteristics of the observed radiation were used to
constrain the properties of the emission region. An isotropic fireball,
as well as a top-hat jet (i.e.~a jet with sharp edges) were ruled out
early on \citep{Kasliwal2017}.  However, it was only with VLA
observations that the presence of a relativistic collimated jet --
suggested by early modeling \citep{Lazzati2018,Ioka2018} and by the
steep radio decay \citep{Lamb2018,Lamb2019}-- was confirmed beyond doubt
\citep{Ghirlanda2019,Mooley2018}, hence establishing the consistency
with a standard, cosmological SGRB observed off-axis.

The production of jets by astrophysical sources, which is an essential
ingredient for both long and short GRBs, is an area of much interest in
astrophysics. In order to understand the mechanisms by which jets are
produced and launched, the first step is the characterization of their
\textit{intrinsic} properties, i.e.~the jets properties as released by
their central engines, before any interaction with the surrounding
material. However, what we observe are the properties of the outflow
when it becomes transparent to radiation, molded by the environment in
which it has propagated. In the case of long GRBs this environment is
the envelope of a massive star  \citep{Macfadyen2001}, while in the case
of SGRBs it is the material expelled in a compact binary merger (e.g.,
\citealt{Rosswog1999,Fernandez2013,Ciolfi2017,Radice2018,Ciolfi2019}).

A model able to compute the SGRB outflow properties resulting from the
jet interaction with the surrounding material was recently developed by
\citet{Lazzati2019}, employing a semi-analytical method calibrated via
numerical simulations (see also \citealt{Salafia2020}). Such model takes
as input the properties of the surrounding material (most importantly
its mass and velocity), those of the jet (namely its asymptotic Lorentz
factor, injection angle, and time delay between the merger and the jet
launching), and the viewing angle, i.e.~the angle between the jet axis
and the line of sight. Here, we apply this model to constrain the
injection parameters of the jet from GW170817. For the properties of the
surrounding material, we refer to the results of general relativistic
magnetohydrodynamics (GRMHD) simulations of BNS mergers performed by
\citet{Ciolfi2017}. We also consider a more general parametric
description as an alternative. For the jet intrinsic properties, we
explore a conservative range for all the relevant parameters.

Constraints for the time interval between merger and jet launching have
been discussed before in the literature, with somewhat controversial
results. Studies based on the need to eject enough material to support a
kilonova \citep{Gill2019} and structure in the jet \citep{Granot2017}
favor a long merger-jet delay of the order of one second. Such delay,
however, requires a coincidence with the propagation time of the jet to
yield a total observed delay of $\sim 1.74$~s. This, and the fact that
the pulse duration of GRB170817A coincides with the total observed delay
favors instead a much shorter merger-jet delay
\citep{Lin2018,Zhang2018,Zhang2019}. Short time delays have also been
suggested by population synthesis calculations of short GRBs 
\citep{Belczinsky2006, Beniamini2020a}.

\begin{table*}[t]
    \caption{Physical quantities and ranges of prior distributions for the input parameters.}
    \centering
    \begin{tabular}{c|c|c|c}
        Symbol & Range & Units & Explanation  \\ \hline \hline
        $E_{\rm{j}}$ & $5\times10^{48} - 2\times10^{50}$ & erg & Total jet energy \\
        $T_{\rm{eng}}$ & $0.1-2.0$ & s & Duration of the engine activity \\
        $L_{\rm{j}}$ & derived & erg/s & Jet luminosity (constant over the engine activity) \\
        $\eta$ & $10-3000$ & $-$ & Asymptotic Lorentz factor of the jet \\
        $\theta_{\rm{j}}$ & $1-45$ & degrees & Initial jet half-opening angle at injection \\
        $\Delta{t}_{\rm{m-j}}$ & $0-1.75$ & s & Time delay between merger and jet launching time \\
        $\theta_{\rm{l.o.s.}}$ & $1-45$ & degrees & Viewing angle with respect to the jet axis \\ \hline
    \end{tabular}
    \label{tab:jetproperties}
\end{table*}

Our paper is organized as follows: Section \ref{sec:methods} describes
the employed methods, based on the model developed by
\citet{Lazzati2019}, as well as the range of values allowed for the
input parameters (for the jet and the surrounding material) and the
observational constraints from GW170817/GRB\,170817A that we enforce.
The results of our study are presented in Section \ref{results}. Then,
we summarize and discuss our results in Section \ref{discussion}.

\section{Methods}
\label{sec:methods}

Our reference scenario is a BNS merger forming a (meta)stable massive NS
remnant which might eventually collapse to a black hole (BH). We assume
that a SGRB jet is launched at a time $\Delta t_{\rm{m-j}}$ after
merger, either by the massive NS or right after BH formation (see, e.g.,
\citealt{Ciolfi2018}). In both cases, a nearly isotropic baryon-loaded
wind from the NS remnant continuously pollutes the surrounding
environment for a time $\Delta t_{\rm{m-j}}$ before the jet is launched.
Our model describes the propagation of the incipient jet across such an
environment and the resulting properties and structure of the final
escaping outflow.\footnote{We note that we are not considering dynamical
(tidal and shock-driven) ejecta from the merger process itself as a
potential obstacle for the jet propagation as these are mostly expelled
at high latitude (i.e.~away from the jet axis). Moreover, this matter is
ejected only within $\sim\!10$\,ms from the merger time and at larger
speed, thus already far away by the time the jet is launched.}
Throughout the manuscript, we will refer to the incipient collimated
outflow from the central engine with high-entropy (and eventually high
Lorentz factor) as ``jet'', to the wide-angle non-relativistic matter
released by the massive NS remnant prior to jet launching as ``wind'',
and to the ultimate structured outflow at large distances resulting from
the jet-wind interaction as ``outflow''.

The analysis that we present is based on the jet-wind interaction model
developed by \cite{Lazzati2019}. By imposing energy conservation and
pressure balance at the jet, cocoon and wind interfaces
\citep{Begelman1989,Matzner2003,Lazzati2005,Morsony2007,Bromberg2011},
they were able to develop a set of semi-analytic equations to compute
the properties of the outflow for any given jet and wind setup. The
underlying assumptions are the following: (i) the jet has initially a
top-hat structure, with uniform properties within a half-opening angle
$\theta_{\rm{j}}$; (ii) the engine turns on at time $\Delta
t_{\rm{m-j}}$ after merger, releasing a constant luminosity $L_{\rm{j}}$
for a time $T_{\rm{eng}}$ and then turning off; (iii) the jet is
characterized by a constant dimensionless entropy $\eta$, which
corresponds to the maximum asymptotic Lorentz factor that the jet
material would attain if the acceleration were complete and
dissipationless.
\begin{figure}[!t]
    \centering
    \includegraphics[width=\columnwidth]{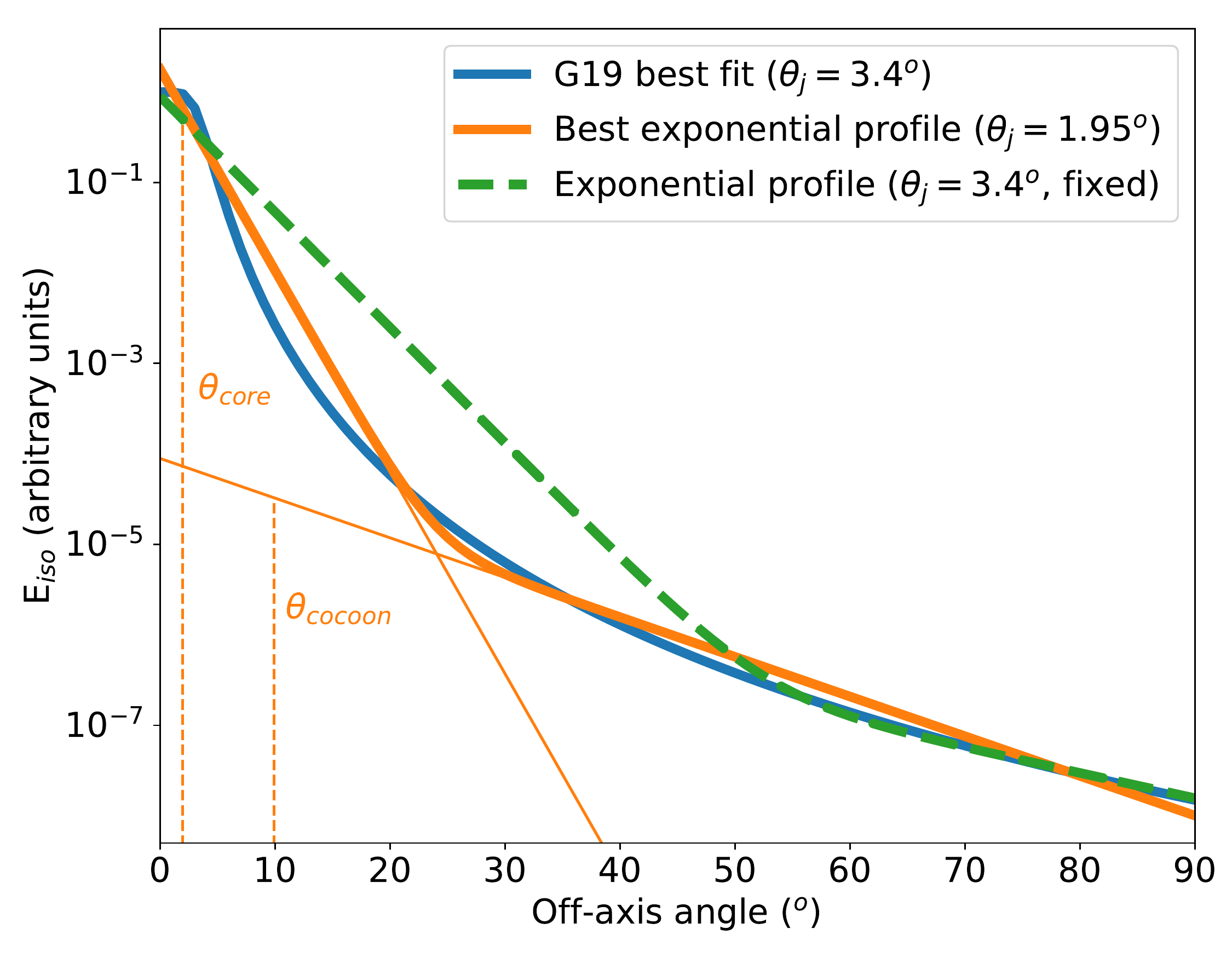}
    \caption{Comparison between the best fit outflow structure from
    \citet[G19 in the legend]{Ghirlanda2019} and the exponential profile
    used in this work. It is found that a scaling factor of 1.75 between
    the core opening angles is necessary for a good match of the angular
    the profiles. Also shown with a dashed green line is the
    exponential profile for a core angle equal to the G19 best fit
    value. Vertical dashed orange lines show the values of
    $\theta_{\rm{core}}$ and $\theta_{\rm{cocoon}}$ for the exponential
    outflow profile}.
    \label{fig:profiles}
\end{figure}

For the wind, we consider two different prescriptions. In the first, we
model the wind following the results of GRMHD simulations of BNS mergers
by \citet{Ciolfi2017}. In particular, we refer to the outcome of their
simulations for two possible equations of state (EOS), APR4
\citep{Akmal1998} and H4 \citep{Glendenning1991}, and for two values of
the mass ratio, $q\!=\!1$ and $q\!=\!0.9$,\footnote{The BNS total mass
in the simulations is fixed and differs by only $\approx\!1\%$ from the
one inferred for GW170817.} labelled as q10 and q09, respectively. For
these different cases, we impose an isotropic wind with constant mass
flow rate matching the value given in Fig.~23 of \citet{Ciolfi2017} and
constant velocity equal to the reported escape velocity, namely
$v_{\rm{w}}\!=\!0.11\,c$, $0.12\,c$, $0.13\,c$, and $0.11\,c$ for the
APR4q09, APR4q10, H4q09, and H4q10 models, respectively. In our second
prescription, the wind is instead parametrized and we consider constant
mass flow rate and velocity spanning a wide range of values, namely
$0.001\le\dot{m}_{\rm{w}}/(M_{\odot}\,\rm{s}^{-1})\le1$ and
$0.05\le{}v_{\rm{w}}/c\le0.25$. In all cases, the wind starts at the
time of merger and persists at least until the engine turns off.
\begin{table*}[t]
\caption{Observational constraints on derived physical quantities}
    \centering
    \begin{tabular}{c|c|c|c}
        Symbol & Range & Units & Description  \\ \hline \hline
        $E_{\rm{iso,l.o.s.}}$ & $3\times10^{47}-2\times10^{50}$ & erg & The outflow isotropic-equivalent energy along the line of sight \\
        $\Gamma_{\rm{l.o.s.}}$ & $1.5-10$ & $-$ & The Lorentz factor of the outflow along the line of sight \\
        $\theta_{\rm{core}}$ & $1.5-4$ & degrees & The half-opening angle of the core of the outflow \\
        $\Delta{t}_{\rm{obs}}$ & $1.5-1.75$ & s & The observed delay between merger time and prompt gamma-ray pulse \\ \hline
    \end{tabular}
    \vspace{0.8cm}
    \label{tab:obsconstraints}
\end{table*}

Our analysis proceeds as follows. A random set of parameter values is
first generated for the system. These are the jet entropy $\eta$, total
emitted energy $E_{\rm j}$, half-opening angle $\theta_{\rm{j}}$,
duration of the engine activity $T_{\rm eng}$, delay time of the jet
launching $\Delta t_{\rm{m-j}}$, and viewing angle with respect to the
jet axis $\theta_{\rm{l.o.s.}}$ (see Table~\ref{tab:jetproperties}). For
the parametrized wind case, the list includes also the mass flow rate
$\dot{m}_{\rm{w}}$ and the wind velocity $v_{\rm{w}}$. All these
parameters are randomly drawn from flat prior distributions within a
range that is either theoretically reasonable or constrained by
observations. We assumed the following priors for the injection
properties:
\begin{itemize}
    \item The jet is launched with an asymptotic Lorentz factor
    $10\le\eta=L_{\rm{j}}/\dot{m}c^2\le3000$. The conservative lower
    limit is set by observational constraints (e.g.,
    \citealt{Ghirlanda2019}), while the upper limit is simply set to a
    rather large value.
    \item The jet total energy is limited to $5\times10^{48}\le
    E_{\rm{j}}/\rm{erg}\le2\times10^{50}$. These values are conservative
    compared to the observational constraints (e.g.,
    \citealt{Fong2015}).
    \item The initial half-opening angle of the jet is limited to
    $1^\circ\le\theta_{\rm{j}}\le45^\circ$. In this case we strove to
    consider a range as large as possible. The lower limit of 1 degree
    is set to avoid a divergence at 0, while the upper limit of
    $45^\circ$ is conservatively larger than any successful jet that has
    been numerically studied 
    \citep{Murguia2014,Murguia2017,Nagakura2014,Lazzati2017,Nakar2018,
    Xie2018,Hamidani2020,Lyutikov2020}.
    \item The delay time between the BNS merger and the jet launching is
    limited to $0\le\Delta t_{\rm{m-j}}\le1.75$\,s. The upper limit in
    this case is set by the observed time delay \citep{LVC2017-GRB}.
    \item The viewing angle is limited to
    $1^\circ\le\theta_{\rm{l.o.s.}}\le45^\circ$. As for the injection
    angle, the lower limit is set to avoid a divergence at 0, while the
    upper limit is larger than the one obtained from both gravitational
    waves and electromagnetic observations
    \citep{LVC2017-GRB,Mooley2018,Ghirlanda2019}.
    \item The duration of the engine activity is limited to
    $0.1\,\rm{s}\le T_{\rm{eng}}\le2$\,s. In this case the lower limit
    is set to avoid a divergence at 0, while the upper limit is chosen
    to be at the traditional separation between long and short gamma-ray
    bursts \citep{Kouveliotou1993}.
\end{itemize}

\begin{figure*}
    \centering
    \includegraphics[width=\textwidth]{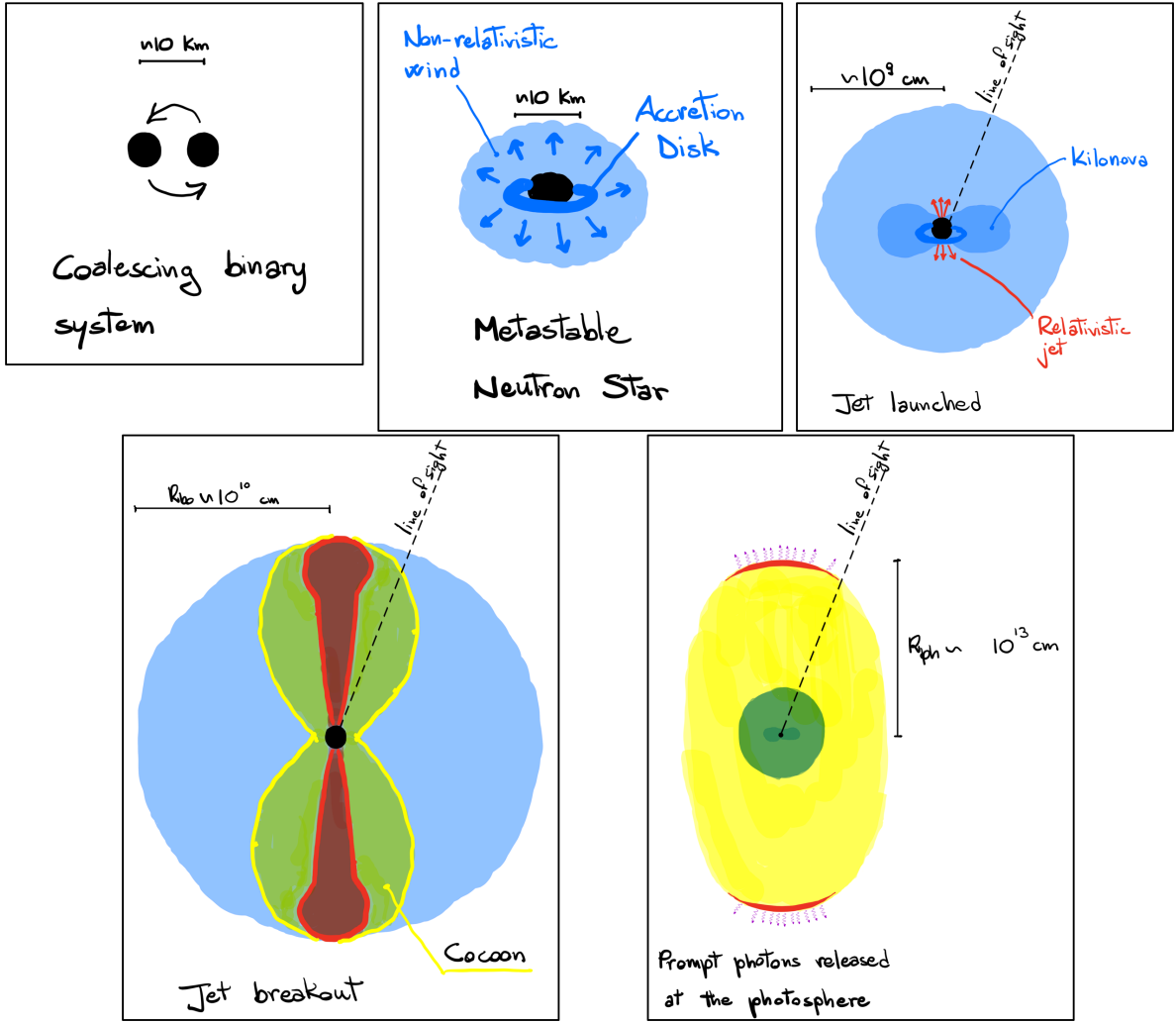}
    \caption{ Cartoon of the various phases of the merger/outflow
    phenomenology, indicating the relevant radii. Numerical values are
    order-of-magnitude estimates, the actual values changing for each
    simulation.}
    \label{fig:cartoon}
\end{figure*}

Once the jet and wind parameters have been drawn, the code computes the
properties of the outflow.\footnote{The jet is launched from a nozzle at
$r_{\rm{0}}=10^{7}$~cm with an initial Lorentz factor of $\Gamma=1$.}
The procedure is repeated for over 100 million random samples. The
resulting outflow properties are then checked against further
observational constraints and only consistent models are retained. The
additional constraints that we enforce are the following (see also
Table~\ref{tab:obsconstraints}):
\begin{itemize}
    \item The isotropic equivalent energy of the outflow in the
    direction of the line of sight has to be within the range
    $3\times10^{47}\le E_{\rm{iso,l.o.s.}}/\rm{erg}\le2\times10^{50}$.
    The lower limit is set by assuming an efficiency of $10\%$ for the
    prompt gamma-ray emission \citep{LVC2017-GRB}. The upper limit is
    obtained by analyzing various best fit models from the literature
    \citep{Alexander2018,Davanzo2018,Lazzati2018,Nakar2018a,Mooley2018,
    Wu2018,Ghirlanda2019,Hotokezaka2019}.
    
    \item The half-opening angle of the core of the outflow (or final
    escaping jet) is limited to 
    $1.3^\circ\le\theta_{\rm{core}}\le4^\circ$. This constraint comes
    exclusively from the modeling of proper motion and spatial extent of
    the radio counterpart \citep{Mooley2018,Ghirlanda2019}. Note that
    both \citet{Mooley2018} and \citet{Ghirlanda2019} use power-law
    outflow models, while here we use a double exponential profile. To
    compensate for such difference, we re-scaled by a factor of 1.75 the
    opening angle values suggested by their analyses. As shown in
    Fig.~\ref{fig:profiles}, this compensation provides a rather good
    match between our angular profiles and theirs.
   
    \item The observed time delay between the merger (or the peak of the
    GW signal) and the gamma-ray detection is constrained to be
    $1.5\,\rm{s}\le\Delta t_{\rm{obs}}\le1.75$\,s and is given by the
    sum of three terms \citep{Zhang2019}:
    \begin{equation}
    \Delta{t}_{\rm{obs}}=\Delta{t}_{\rm{m-j}}+\frac{R_{\rm{bo}}}{c}
    \frac{1-\beta_{\rm{jh}}}{\beta_{\rm{jh}}}
    +\frac{R_{\rm{ph, \, l.o.s.}}-R_{\rm{bo}}}{c} 
    \frac{1-\beta_{\rm{l.o.s.}}}{\beta_{\rm{l.o.s.}}}\,,
    \label{eq:dtobs}
    \end{equation}
    where $R_{\rm{bo}}$ is the radius at which the jet breaks out of the
    wind, $\beta_{\rm{jh}}$ is the speed of the head of the jet inside
    the wind in units of $c$, $R_{\rm{ph,\,l.o.s.}}$ is the photospheric
    radius of the outflow, and $\beta_{\rm{l.o.s.}}$ is its velocity in
    units of $c$, both measured along the line of sight of the
    observation. Figure~\ref{fig:cartoon} shows the location of the
    various radii throughout the evolution of the merger and subsequent
    outflow. Here we have considered a fairly wide interval, down to
    1.5\,s, to take into account the fact that the beginning of the
    gamma-ray emission may have been misidentified if initially below
    the the background.
    
    \item The initial Lorentz factor of the material moving along the
    line of sight is within the interval
    $1\le\Gamma_{\rm{l.o.s.}}\le10$. This is a  conservative constraint
    obtained from combining various afterglow models
    \citep{Alexander2018,Davanzo2018,Lazzati2018,Mooley2018,Wu2018,
    Ghirlanda2019,Hotokezaka2019,Beniamini2020}.
\end{itemize}

\subsection{Calculation of the photospheric radius}

A critical piece of information for constraining the observed time delay
is the calculation of the location of the photosphere (see
Eq.~\ref{eq:dtobs}). Calculations of the photospheric radius in
gamma-ray burst outflows have been commonly performed either in the
approximation of a thin shell or of an infinite wind (e.g.,
\citealt{Meszaros2000,Daigne2002}). A large Lorentz factor for which
$(1-\beta)\simeq1/2\Gamma^2$ has also been assumed. In the case of
off-axis outflows, all approximations should be relaxed, since
relatively slow outflows in thick -- but not infinite -- shells are
relevant. In addition, it has been customary to assume a neutron free
fireball in past GRB literature, for which
$Y_{\rm{e}}\equiv\frac{n_{\rm{p}}}{n_{\rm{p}}+n_{\rm{n}}}=1$. Here,
$n_{\rm{p}}$ and $n_{\rm{n}}$ are the proton and neutron densities,
respectively, and we generalize the equations for the photospheric
radius to the case of an outflow with $Y_{\rm{e}}\le1$. We assume our
fiducial electron fraction to be $Y_{\rm{e}}=0.5$ or lower, as expected
for most GRB engines \citep{Beloborodov2003}, but quote also results for
$Y_{\rm{e}}=1$.

Let us consider a photon that is at the back of the outflow. If its
location corresponds to the photospheric radius, then the photon has
probability 1/2 of undergoing a scattering before leaving the flow at
the front. We can therefore write a condition on the opacity such that
\begin{equation}
    \tau=\frac{2}{3}=\int_{R_{\rm{ph}}}^{R_{\rm{ph}}+\Delta} n_{\rm{e}}\sigma_T
    \left(1-\beta\cos(\theta_{\gamma{e}})\right) \; dr\,,
\end{equation}
where $R_{\rm{ph}}+\Delta$ is the outer radius of the outflow at the
time at which the photon leaves the outflow, $n_{\rm{e}}$ is the
fireball's electron number density in the observer frame, and
$\theta_{\gamma{e}}$ is the angle between the photon's and the outflow's
velocity vectors. Assuming $\theta_{\gamma{e}}\sim1/\Gamma$, we have
\begin{equation}
    \frac{2}{3}=\frac{L_{\rm{iso, l.o.s.}} Y_{\rm{e}}\sigma_T\left(1-\beta\right)}{4\pi m_p c^3 \eta}
    \int_{R_{\rm{ph}}}^{R_{\rm{ph}}+\frac{cT_{\rm{eng}}}{1-\beta}} 
     \frac{dr}{r^2}\,,
\end{equation}
where $\sigma_T$ is the Thomson cross section, and we have used
\begin{equation}
    n_{\rm{e}}=\frac{L_{\rm{iso, l.o.s.}}Y_{\rm{e}}}{4\pi r^2  m_{\rm{p}} c^3 \eta}\,.
\end{equation}
We have also assumed that the fireball is fully accelerated by the time
it reaches the photospheric radius, which is reasonable for a low
Lorentz factor outflow. Here we have used the subscript $_{\rm{
l.o.s.}}$ to remind the reader that the calculated photospheric radius
is for material moving along the line of sight to the observer.

\begin{figure}
    \centering
    \includegraphics[width=\columnwidth]{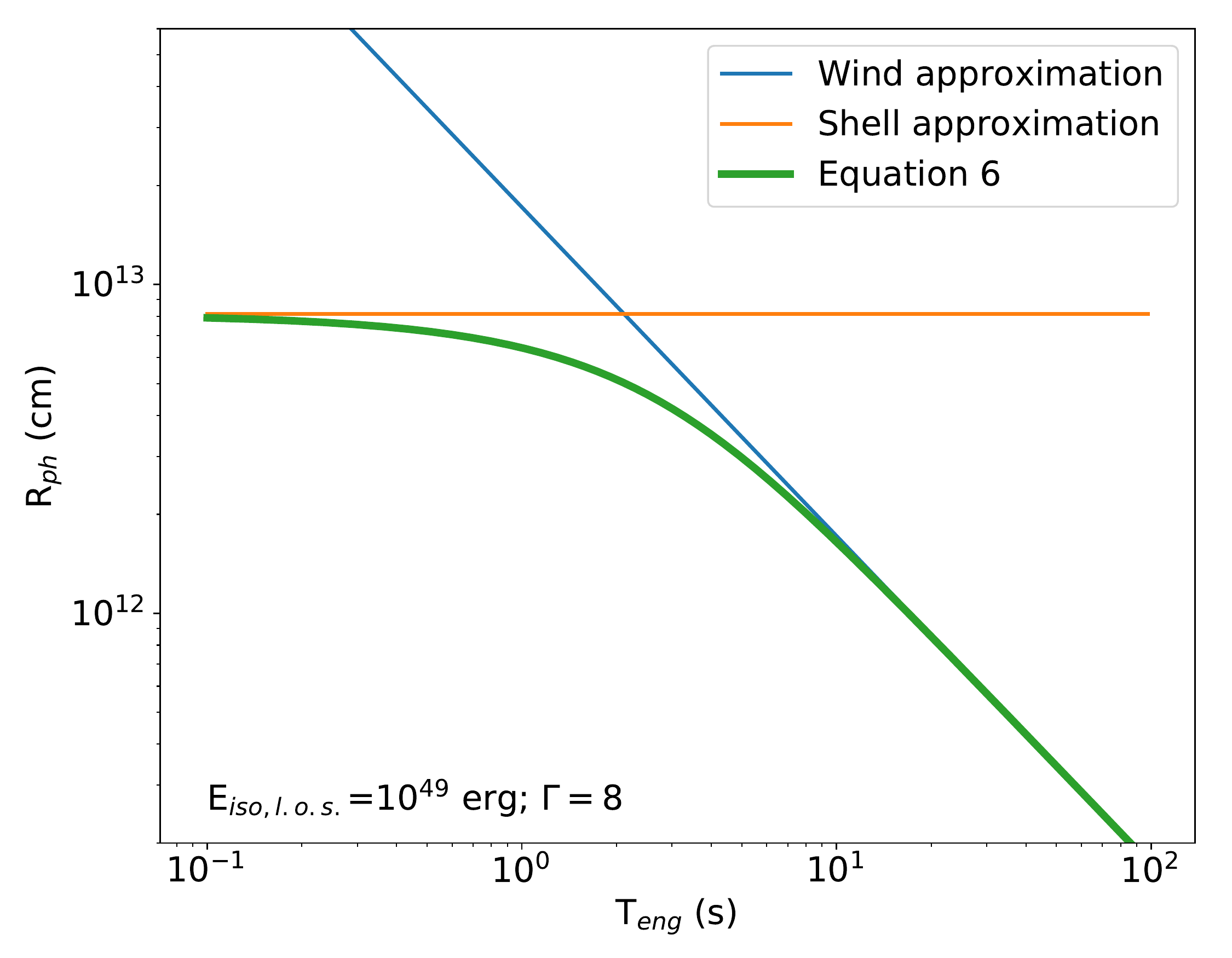}
    \caption{Comparison between the solution for the photospheric radius
    given in Eq.~(\ref{eq:rph}) and the approximations for an infinite
    wind and a thin shell.}
    \label{fig:rph}
\end{figure}

Performing a trivial integration we obtain
\begin{equation}
    \frac{2}{3}=\frac{L_{\rm{iso, l.o.s.}}Y_{\rm{e}}\sigma_T\left(1-\beta\right)}{4\pi m_p c^3 \eta}
    \left(\frac{1}{R_{\rm{ph}}}-\frac{1}{R_{\rm{ph}}+\frac{cT_{\rm{eng}}}{1-\beta}}
    \right)\,,
\end{equation}
which is solved to yield
\begin{equation}
    R_{\rm{ph}}=
    \frac{\sqrt{\left(\frac{cT_{\rm{eng}}}{1-\beta}\right)^2
    +\frac{3}{2\pi}\frac{L_{\rm{iso, l.o.s.}}Y_{\rm{e}}\sigma_T T_{\rm{eng}}}{m_p c^2 \eta}}-
    \frac{cT_{\rm{eng}}}{1-\beta}
    }{2}\,.
    \label{eq:rph}
\end{equation}
Note that the latter equation is valid for any shell thickness, and that
it has the correct asymptotic behavior for a high-Lorentz factor wind
case, for which $(1-\beta)=1/2\eta^2$:
\begin{equation}
    \lim_{T_{\rm{eng}}\to\infty;\, \eta\to\infty} R_{\rm{ph}} = \frac{3}{16\pi}
    \frac{L_{\rm{iso, l.o.s.}}Y_{\rm{e}}\sigma_T}{\eta^3m_pc^3}\,.
    \label{eq:rphw}
\end{equation}
In the opposite extreme of a thin fireball, we obtain\footnote{In this
case we note that the result differs by a factor $\eta$ from the
equation that was previously derived, e.g., in \citet{Lazzati2017a}
since that derivation did not consider the expansion of the outer edge
of the fireball while the photon crosses it.}
\begin{equation}
    \lim_{T_{\rm{eng}}\to 0;\, \eta\to\infty} R_{\rm{ph}} = \sqrt{
    \frac{3E_{\rm{iso, l.o.s.}}Y_{\rm{e}}\sigma_T}{2\pi \eta m_p c^2}}\,.
    \label{eq:rphs}
\end{equation}
Figure~\ref{fig:rph} shows, for an outflow with properties similar to
those revealed by GRB\,170817A along the line of sight, how the result
of Eq.~(\ref{eq:rph}) depends on the engine duration $T_{\rm{eng}}$.
Also shown are the two limiting cases of wind and shell approximations,
correctly recovered. In all the calculations of this paper, we use the
more general Eq.~(\ref{eq:rph}).
\begin{figure*}
    \centering
    \includegraphics[width=0.9\textwidth]{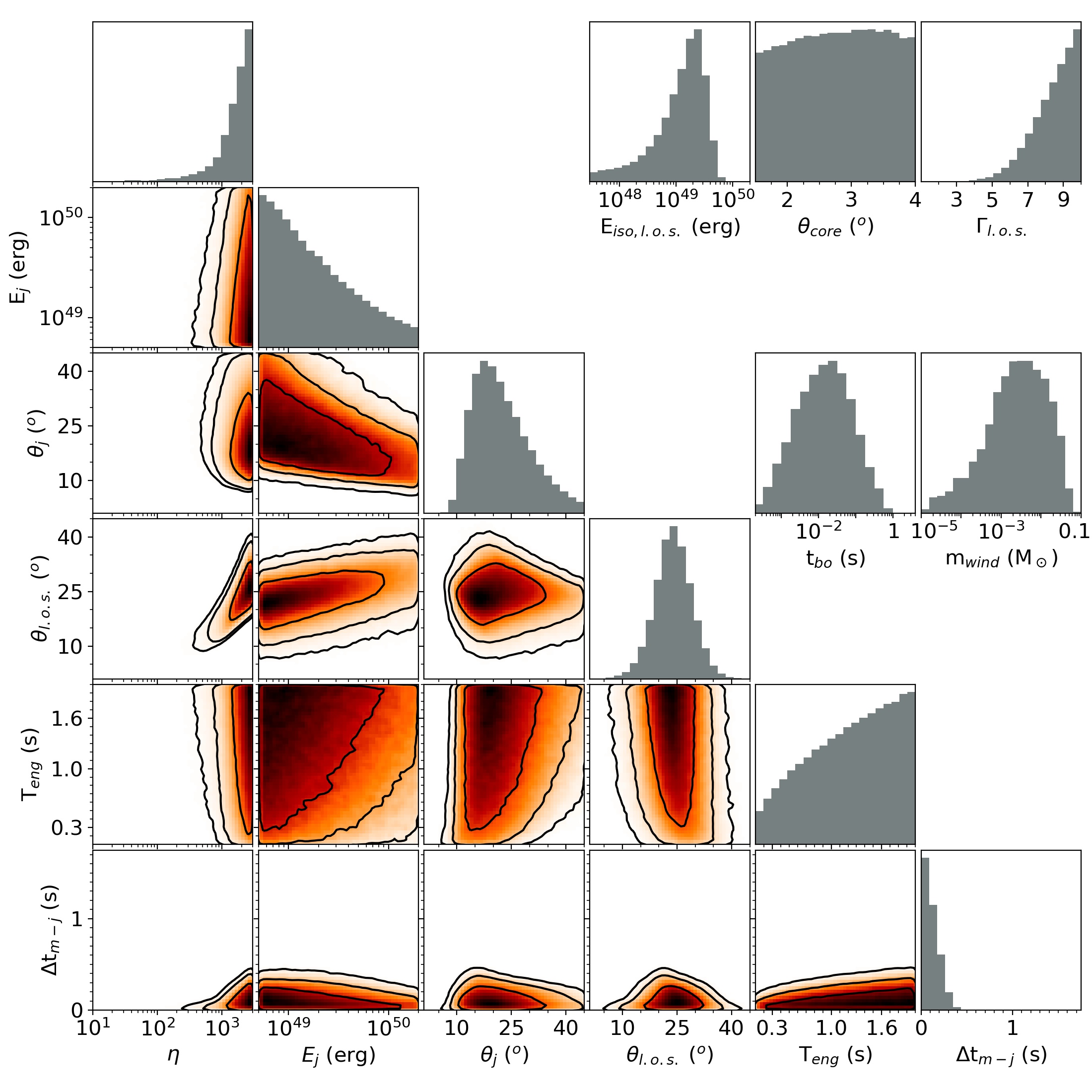}
    \caption{Corner diagram for the six parameters of the model adopting
    the wind prescriptions inspired by GRMHD simulations by
    \citet{Ciolfi2017}. The results for the four different combinations
    of EOS and mass ratios are merged together. Solid contour lines show
    the $1\sigma$, $2\sigma$, and $3\sigma$ confidence regions. In
    addition, probability distributions for the four derived parameters
    and for the breakout time $t_{\rm{bo}}$ are shown as
    histograms in the upper right corner. Note that these show the ratio
    of the posterior over the prior distributions.}
    \label{fig:170817_corner}
\end{figure*}
\begin{figure*}[ht]
    \centering
    \includegraphics[width=0.8\textwidth]{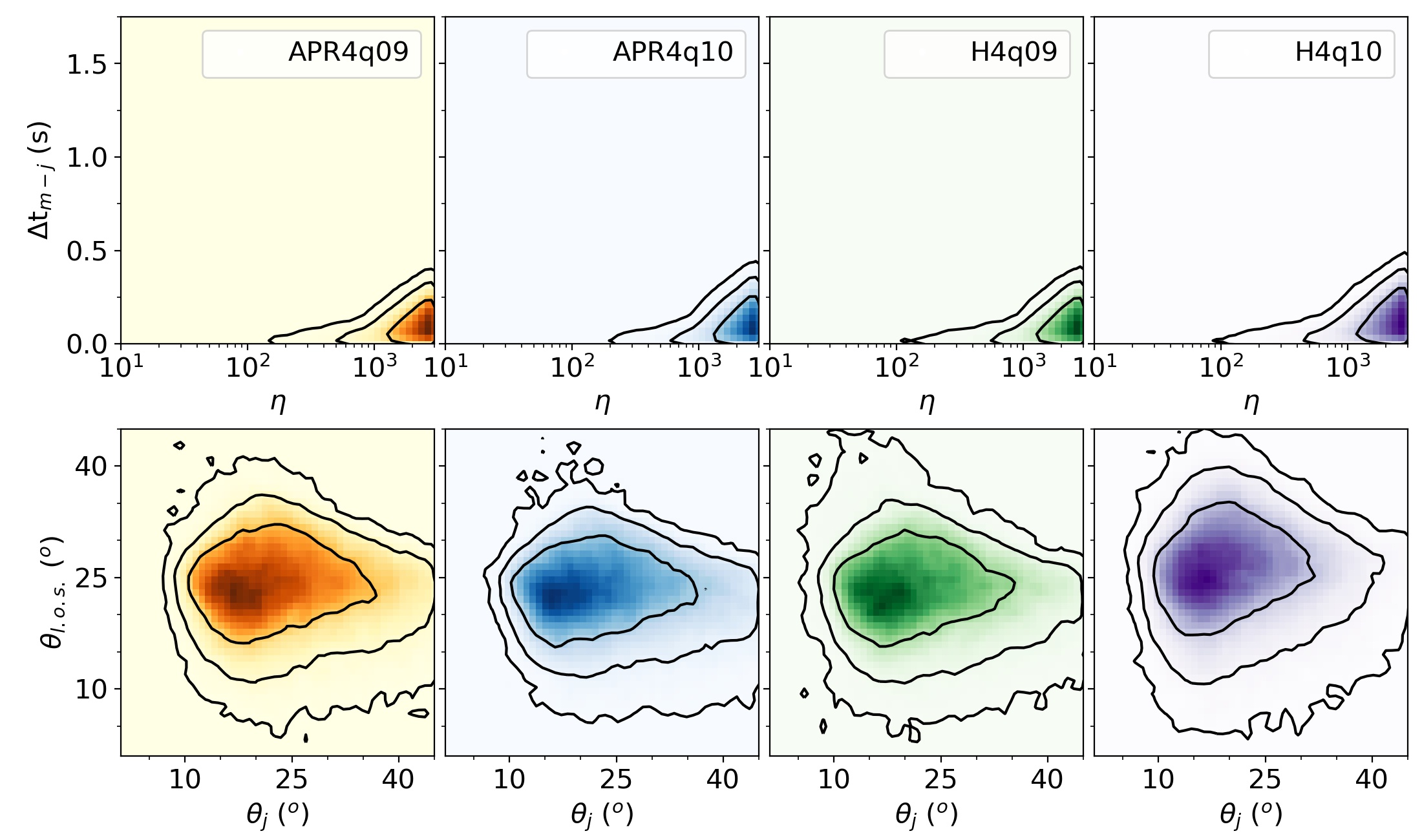}
    \caption{Correlation plots of the two best constrained parameter
    pairs. The top row shows the $\Delta t_{\rm{m-j}}-\eta$ plane, while
    the bottom row shows the $\theta_{\rm{l.o.s.}}-\theta_{\rm{j}}$
    plane. For both parameter pairs, four panels are shown (left to
    right), corresponding to the four different EOS-mass ratio
    combinations considered. Solid contour lines show the $1\sigma$,
    $2\sigma$, and $3\sigma$ confidence regions.}
    \label{fig:170817_simulae}
\end{figure*}
\begin{figure*}
    \centering
    \includegraphics[width=0.8\textwidth]{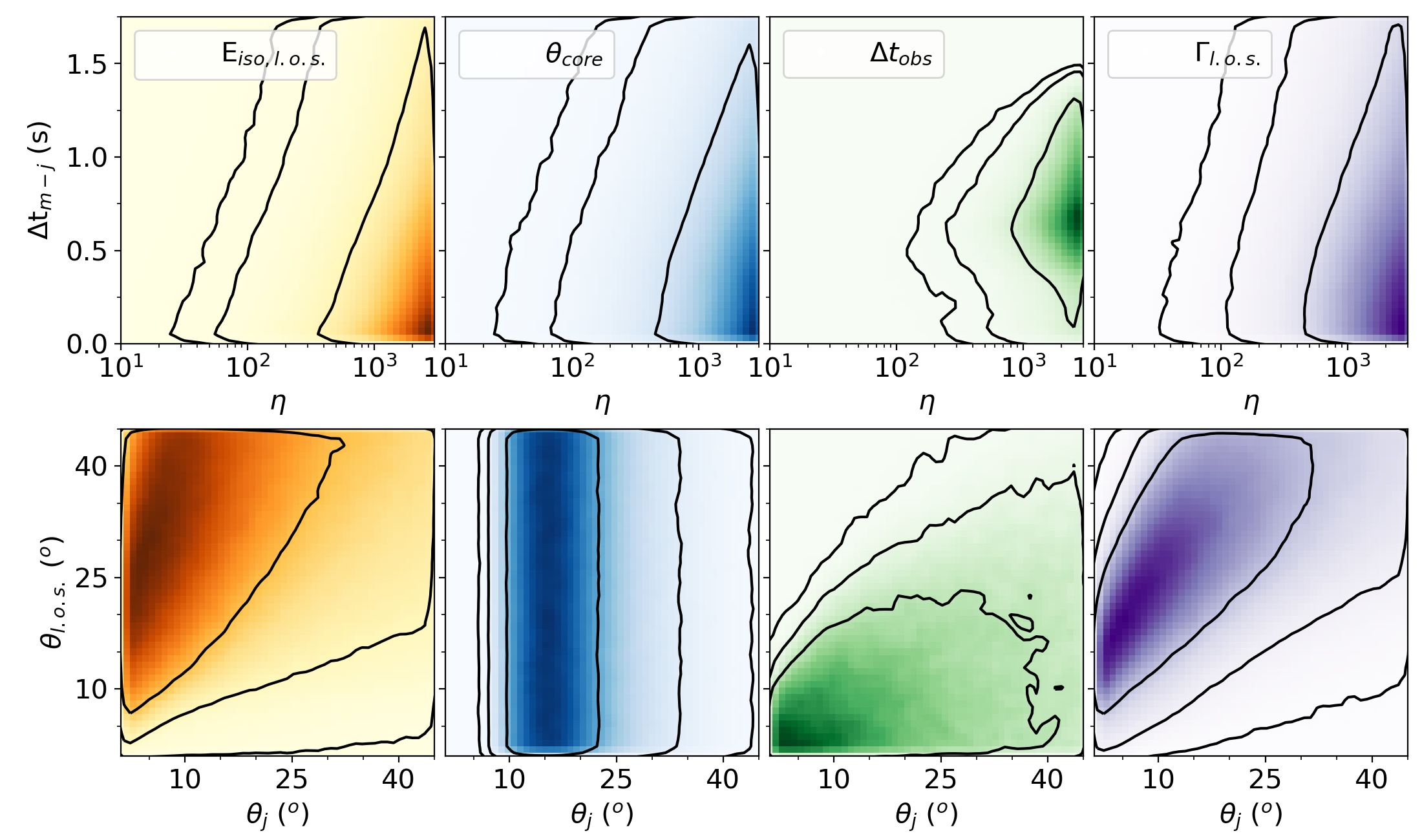}
    \caption{Correlation plots of the two best constrained parameter
    pairs. The top row shows the $\Delta t_{\rm{m-j}}-\eta$ plane, while
    the bottom row shows the $\theta_{\rm{l.o.s.}}-\theta_{\rm{j}}$
    plane. For both parameter pairs, four panels are shown (left to
    right) with the results obtained by imposing only one observational
    constrain at a time (see text). Solid contour lines show the
    $1\sigma$, $2\sigma$, and $3\sigma$ confidence regions.}
    \label{fig:170817_constraints}
\end{figure*}
\begin{figure*}[!t]
    \centering
    \includegraphics[width=0.9\textwidth]{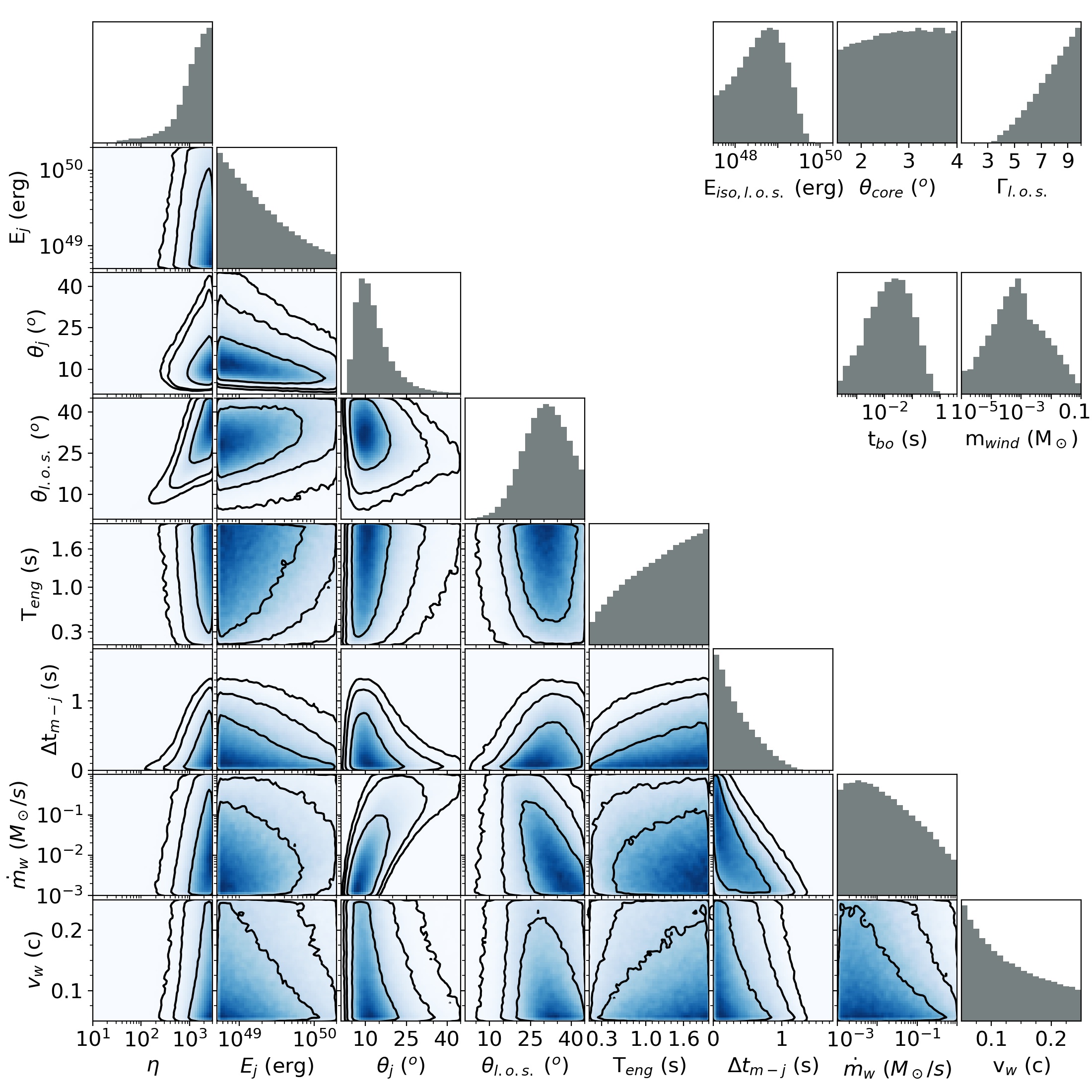}
    \caption{Analogous to Figure~\ref{fig:170817_corner} but for the
    case of a parametrized wind. In this case, the model has two
    additional parameters, i.e.~the mass flow rate and the velocity of
    the wind.}
    \label{fig:free_mdot_corner}
\end{figure*}

\section{Results}
\label{results}

\begin{table*}[t]
    \caption{Results for the four most constrained parameters: the
    merger-jet delay, the asymptotic Lorentz factor of the jet, the
    viewing angle, and the injection angle. Quoted uncertainties are at
    the $1\sigma$ level, while upper and lower limits are $3\sigma$. We
    highlight in bold the results for our baseline model for both the
    simulation-inspired wind and the parametric wind cases.}
    \centering
    \begin{tabular}{l|c|c|c|c}
        Model & $\Delta{t_{\rm{m-j}}}$ (s) & $\eta$ & $\theta_{\rm{l.o.s}}$ ($^{\rm{o}}$) & $\theta_{\rm{j}}$ ($^{\rm{o}}$) \\ \hline \hline 
        
         {\bf Simulations; baseline ($\mathbf{Y_{\rm{e}}=0.5}$; $\mathbf{\Gamma_{\rm{l.o.s.}}\le10}$; $\mathbf{m_{\rm{w}}}$ unconstrained)}  &
        $\mathbf{<0.36}$ & $\mathbf{>240}$ & $\mathbf{23.5^{+5.5}_{-4.5}}$ & $\mathbf{17.9^{+12.6}_{-3.2}}$\\

        Simulations; $\Gamma_{\rm{l.o.s.}}\le7$ &
        $<0.18$ & $>240$ & $24^{+6.9}_{-3.5}$ & $18.4^{+12.5}_{-3.1}$\\

        Simulations; $m_{\rm{w}}\ge10^{-2}$ &
        $<0.37$ & $>390$ & $23.6^{+4.8}_{-4.5}$ & $17.3^{+13.4}_{-2.5}$\\

        Simulations; $\Gamma_{\rm{l.o.s.}}\le7$; $m_{\rm{w}}\ge10^{-2}$ &
        $<0.17$ & $>250$ & $24.1^{+6.7}_{-3.6}$ & $19.3^{+11.9}_{-3.9}$\\ \hline

       Simulations; $Y_{\rm{e}}=1.0$ & $<0.27$ & $>260$ & $22.0^{+5.9}_{-}3.3$ & $18.1^{+13.4}_{-3.1}$ \\
        
        Simulations; $Y_{\rm{e}}=0.2$ & $<0.51$ & $>170$ & $25.1^{+5.0}_{-6.0}$ & $15.8^{+13.2}_{-1.9}$ \\ \hline \hline
      
        {\bf Parametric; baseline ($\mathbf{Y_{\rm{e}}=0.5}$; $\mathbf{\Gamma_{\rm{l.o.s.}}\le10}$; $\mathbf{m_{\rm{w}}}$ unconstrained)}  &
        $\mathbf{<1.1}$ & $\mathbf{>150}$ & $\mathbf{30.3^{+8.5}_{-8.0}}$ & $\mathbf{10.2^{+8.8}_{-3.0}}$ \\

        Parametric; $\Gamma_{\rm{l.o.s.}}\le7$ &
        $<0.87$ & $>180$ & $34.4^{+6.4}_{-8.6}$ & $9.2^{+9.7}_{-1.8}$ \\

        Parametric; $m_{\rm{w}}\ge10^{-2}$ &
        $<0.87$ & $>420$ & $27.5^{+6.0}_{-7.1}$ & $16.2^{+11.3}_{-3.2}$ \\

        Parametric; $\Gamma_{\rm{l.o.s.}}\le7$; $m_{\rm{w}}\ge10^{-2}$ &
        $<0.57$ & $>800$ & $30.7^{+6.2}_{-6.8}$ & $16.3^{+13.8}_{-1.2}$ \\ \hline
        
        Parametric; $Y_{\rm{e}}=1.0$ & $<1.0$ & $>170$ &
        $32.3^{+6.4}_{-9.5}$ & $9.6^{+9.0}_{-2.5}$ \\
 
        Parametric; $Y_{\rm{e}}=0.2$ & $<1.2$ & $>130$ & 
        $30.5^{+8.3}_{-8.8}$ & $10.8^{+8.6}_{-3.6}$ \\ \hline \hline
        
    \end{tabular}
    \vspace{0.8cm}
    \label{tab:results}
\end{table*}

The results of the analysis are best shown through corner plots, where
each of the model parameters is plotted versus the other ones. In the
corner plot figures, the colored panels show the density map of models
that satisfy the observational constraints, while the solid lines mark
the areas of $1\sigma$, $2\sigma$, and $3\sigma$ statistical
significance level. Histograms on the diagonal show the posterior
probability distribution for each parameter marginalized over the
others. Finally, histograms in the upper right part of the figures show
the posterior probability distribution for the observational quantities
of interest.

In Figure~\ref{fig:170817_corner}, we report the outcome for the wind
properties inspired by the GRMHD simulations of \citet{Ciolfi2017}.
Here, we are combining together the four different cases APR4q09,
APR4q10, H4q09, and H4q10, and we show the outcome of the simulations
for our baseline case with $Y_{\rm{e}}=0.5$. We found that some of the
parameters are well constrained. To begin with, the viewing angle, which
was not directly constrained in our procedure, is constrained to
$\theta_{\rm{l.o.s.}}=23.5^{+5.5}_{-4.5}$ degrees (all quoted
uncertainties are at the $1\sigma$ statistical significance level,
unless stated otherwise), a value that is in good agreement with the
estimates based on high-resolution radio imaging
\citep{Mooley2018,Ghirlanda2019}. 

Parameters for which we cannot obtain direct limits from observations
and which are also well constrained are $\theta_{\rm{j}}$, $\eta$, and
$\Delta{t}_{\rm{m-j}}$. The injection half-opening angle, never measured
for long or short GRBs, is found to be
$\theta_{\rm{j}}=17.9^{+12.6}_{-3.2}$ degrees. Additionally, we obtained
a lower limit for the dimensionless jet entropy (i.e.~the maximum
attainable Lorentz factor) as $\eta>240$ at the $3\sigma$ level.
Finally, we found that the delay time between the merger and the
injection of the jet is bound to be rather small:
$\Delta{t}_{\rm{m-j}}<0.36$\,s. These values are also reported in
Table~\ref{tab:results}, which further shows how such constraints change
by considering different electron fractions ($Y_{\rm{e}}=1.0$ and $0.2$)
and stricter constraints on $\Gamma_{\rm{l.o.s.}}$ and/or the total wind
mass $m_{\rm{w}}$.

For the remaining parameters, our results favor jet energies at the
lower edge of the simulated values ($E_{\rm{j}}\sim5\times10^{48}$~erg),
engine activity duration  $T_{\rm{eng}}\sim~2$\,s, line-of-sight Lorentz
factor $\Gamma_{\rm{l.o.s.}}\gtrsim6$, isotropic equivalent outflow
energy along the line of sight $E_{\rm{iso,\,
l.o.s.}}\sim2\times10^{49}$~erg, and a total mass of the wind in the
range $m_{\rm{wind}}\sim10^{-3}-10^{-2}$\,$M_\odot$. We note that the
finding on the outflow energy is in general agreement with previous
constraints from the afterglow modeling.

To check whether our results are sensitive to the different EOS and/or
mass ratios under consideration, we show in
Figure~\ref{fig:170817_simulae} the two panels
$\Delta{t}_{\rm{m-j}}$\,vs.\,$\eta$ and
$\theta_{\rm{l.o.s.}}$\,vs.\,$\theta_{\rm{j}}$, corresponding to the
most constrained parameters from Figure~\ref{fig:170817_corner}, now
separating the four cases. We find that the method is not able to
distinguish among the four, with only a marginal difference in the
$\theta_{\rm{l.o.s.}}$\,vs.\,$\theta_{\rm{j}}$ panel for the H4q10 case
(rightmost lower panel). This degeneracy reflects the fact that the mass
flow rates and velocities are rather similar despite the different $q$
and EOS.

In Figure~\ref{fig:170817_constraints}, we select the same two panels
from Figure~\ref{fig:170817_corner}, but in this case we show how the
result changes by imposing only one of the four observational
constraints at a time. The lower limit on $\eta$ is always reproduced
independently from which constraint is imposed, while for the other
parameters the outcome is significantly affected by the specific choice.
Interestingly, all constraints are consistent with each other at the
$1\sigma$ level, since the $1\sigma$ contours have a non-null
intersection.

We now turn to consider the results obtained with a parametrized wind,
i.e.~allowing for any value of the mass flow rate and wind velocity
within the plausible ranges
$0.001\le\dot{m}_{\rm{w}}/(M_{\odot}\,\rm{s}^{-1})\le1$ and
$0.05\le{}v_{\rm{w}}/c\le0.25$. The outcome, shown in
Figure~\ref{fig:free_mdot_corner}, is qualitatively similar to the
previous case (cf.~Fig.~\ref{fig:170817_corner}), with the viewing angle
and the initial jet half-opening angle well constrained, a lower limit
on the jet dimensionless entropy, and an upper limit on the time
interval between merger and jet launching. 

At a quantitative level, however, some differences emerge. The viewing
and jet angles are constrained to different values, namely
$\theta_{\rm{l.o.s.}}=30.3^{+8.5}_{-8.0}$ degrees and
$\theta_{\rm{j}}=10.2^{+8.8}_{-3.0}$ degrees, which remain nonetheless
consistent within the $1\sigma$ range. The constraints on the
dimensionless entropy and on the time delay are less stringent:
$\eta>150$ and $\Delta{t}_{\rm{m-j}}<1.1$\,s. These variations are
brought about by winds which tend to have smaller velocities and smaller
total masses compared to the values suggested by the GRMHD simulations
of \citet{Ciolfi2017} (cfr. Figure~\ref{fig:free_mdot_corner} and
Table~\ref{tab:results}).

\section{Discussion and conclusions}
\label{discussion}

In this work, we have studied the key properties of the SGRB jet that
was launched by the remnant of the BNS merger event GW170817
\citep{LVC2017-BNS} and that eventually powered the gamma-ray signal
GRB\,170817A \citep{LVC2017-GRB,Goldstein2017,Savchenko2017}. We
employed the semi-analytic model for the jet-wind interaction developed
by \citet{Lazzati2019} to obtain the properties of the escaping outflow
depending on (i) the properties of the jet at the initial injection from
the central engine and (ii) the properties of the massive baryon-loaded
wind expelled beforehand by the NS remnant and acting as an obstacle for
the propagation of the jet itself. By exploring the plausible parameter
ranges with over 100 million random samples and then selecting only
cases with an outcome consistent with four main observational
constraints (see Section~\ref{sec:methods}), we were able to obtain
posterior distributions for the entire parameter set, and hence
indications on their most favourable values.

For the wind properties, we assumed an isotropic flow expelled from the
time of merger to the time of jet launching with constant mass flow rate
and velocity. In our first analysis, the values of the latter were
chosen in accordance to the results of GRMHD BNS merger simulations by
\citet{Ciolfi2017}, referring to BNSs with two different EOS and two
different mass ratios. Then, we considered a more general parametrized
wind and explored a wide range of mass flow rates and velocities.

For the analysis inspired by GRMHD simulations, we found an initial
half-opening angle of the jet of $\theta_{\rm{j}}=17.9^{+12.6}_{-3.2}$
degrees (at $1\sigma$ level) and a robust $3\sigma$ lower limit on the
dimensionless entropy $\eta=L/\dot{m}c^2>240$. We remark that
constraints on these intrinsic jet  properties are of particular
interest, as they cannot be directly obtained from the observations. The
rather large lower limit for the injection entropy suggests a low baryon
loading, as in the case of electromagnetically driven acceleration
mechanisms \citep{Meszaros1997,Drenkhahn2002,Metzger2011}.

In addition, we obtained an upper limit  on the time delay between the
merger and the jet launching:  $\Delta{t}_{\rm{m-j}}<0.36$\,s at the
$3\sigma$ level.\footnote{Here we are assuming a fiducial electron
fraction of $Y_{\rm{e}}=0.5$ within the fireball. For lower values, the
photospheric radius would also be reduced, changing the constraint on
the time delay $\Delta{t}_{\rm{m-j}}$. Even allowing for a quite extreme
$Y_{\rm{e}}=0.2$, however, the upper limit remains rather small
$\Delta{t}_{\rm{m-j}}\lesssim0.51$\,s (i.e.~about a factor $\sqrt{2.5}$
larger, as expected).} This limit would imply that most of the observed
delay ($\approx1.74$\,s) is due to the outflow breaking out of the wind
and its subsequent propagation until the photospheric radius is reached
(along the line of sight), in agreement with the idea that the
similarity between the gamma-ray pulse duration and the total observed
delay is not a simple coincidence \citep{Zhang2018,Lin2018}. This
is also in agreement with populations studies on short GRBs
\citep{Beniamini2020a}. Such a result is likely influenced by the
fairly large prompt emission energetics and, at the same time, by the
fact that the Lorentz factor of the emerging outflow along the line of
sight could not be too large to account for the late onset of the
afterglow emission \citep{Troja2017,Hallinan2017}. These two features,
when taken together, imply that the fireball carried a significant
number of baryons, therefore pushing the photosphere to relatively large
radii. We note that the change from a jet released with high $\eta$
value to an outflow with a significant rest-mass component requires
baryon loading during the interaction of the jet with the wind
material. Since the photosphere location is of such importance for
estimating the propagation delay, we have derived in this paper a
formula for the photospheric radius that relaxes the two commonly used
approximations of either an infinite wind or a thin shell (see
Equation~\ref{eq:rph}).

The above upper limit $\Delta{t}_{\rm{m-j}}\lesssim0.4$\,s has
potentially important implications. In particular, under the assumption
that the central engine launching the jet was a newly-formed BH, as
currently favoured by GRMHD BNS merger simulations
(\citealt{Ruiz2016,Ciolfi2020a}; see \citealt{Ciolfi2020b} for a recent
review), this constraint would imply a NS remnant lifetime $\lesssim
0.4$\,s. In turn, this would help in further constraining the NS EOS, as
well as physical models of the kilonova that accompanied the August 2017
event.

By looking at the other parameters, we note that the total jet energy
and the engine duration are found in general agreement with the
observations (see, e.g., \citealt{Ghirlanda2019} and
\citealt{LVC2017-GRB}, respectively), while the indication on the
Lorentz factor along the line of sight, $\Gamma_{\rm{l.o.s.}}\gtrsim6$,
is at the higher end of (but still consistent with) the range of
available estimates, for which $\Gamma_{\rm{l.o.s.}}$ should not be
larger than $\approx 7$ (e.g., \citealt{Beniamini2020}). The viewing
angle is constrained rather well and is also consistent (within the
$1\sigma$ range) with the latest radio observations
\citep{Mooley2018,Ghirlanda2019}. Finally, the favoured range for the
total mass in the wind is $m_{\rm{wind}}\sim 10^{-3}-10^{-2}\,M_\odot$.
We note that this is only marginally consistent with a scenario in which
(i) the jet was launched after the collapse to a BH \citep{Ciolfi2020a}
and (ii) the wind from the NS remnant is what mainly powered the early
``blue'' component of the associated kilonova (as assumed, e.g., in
\citealt{Gill2019}); indeed, such a scenario would require a mass as
high as $\sim10^{-2}\,M_\odot$ for the unbound portion of the wind
material (e.g., \citealt{Villar2017}).

For completeness, we also checked how the constraints change by imposing
$\Gamma_{\rm{l.o.s.}}\leq7$ (as in \citealt{Beniamini2020}) and/or
$m_{\rm{wind}}\geq 10^{-2}\,M_\odot$ (to better accomodate the
hypothesis of the blue kilonova being powered by the NS remnant wind and
the jet being launched after the collapse to a BH). The additional
condition on $\Gamma_{\rm{l.o.s.}}$ has the interesting effect of
further reducing the upper limit on $\Delta{t}_{\rm{m-j}}$ by a factor
around 2, while the other results are poorly affected. The additional
condition on $m_{\rm{wind}}$ does not show a significant effect on
$\Delta{t}_{\rm{m-j}}$, but makes the lower limit on $\eta$ more
stringent (although this effect disappears when both the additional
conditions are applied).

The analysis based on a parametrized wind confirmed the above overall
picture, although with some quantitative differences. Not surprisingly,
we found that the derived constraints are relaxed once we allow for a
broader range of mass flow rates and wind velocities, especially if we
consider a very low electron fraction. The merger-jet time delay, in
particular, is constrained to $\Delta{t}_{\rm{m-j}}<1.1$\,s (at
$3\sigma$), which is less restrictive. We also note that in this case
small wind velocities (lower than $0.1\,c$) appear to be favoured, as
well as total wind masses no larger than
${\rm{few}}\times10^{-3}\,M_\odot$. Finally, this analysis favours a
viewing angle of $\theta_{\rm{l.o.s.}}=30.3^{+8.5}_{-8.0}$ degrees that
is somewhat larger than what estimated from high resolution radio
imaging \citep{Mooley2018,Ghirlanda2019}, causing some strain with the
observations. In this case, the additional conditions on
$\Gamma_{\rm{l.o.s.}}$ and $m_{\rm{wind}}$ lower significantly the upper
limit on $\Delta{t}_{\rm{m-j}}$, substantially enlarge the lower limit
on $\eta$, and also increase $\theta_{\rm{j}}$ up to values similar to
the simulation-inspired wind case.

As a general note of caution, we remark that in this work we assumed
constant mass flow rates and velocities for the baryon-loaded wind
produced by the NS remnant. This simplifying assumption may have
relevant effects on the outcome of our analysis. Relaxing this
assumption and employing time-evolving wind properties (possibly
motivated by BNS merger simulation results) will be the subject of
future investigation.

While our approach can be further refined, the present study shows its
potential. In particular, the possibility of inferring the intrinsic jet
properties at the time the jet itself is launched by the central engine
can provide a valuable input for the investigation of jet launching
mechanisms via numerical simulations. We also stress that here we
applied the model to the case of GW170817/GRB\,170817A, but our method
is general and can be readily applied to any other SGRB observed in the
future.

\acknowledgements DL acknowledges support from NASA grants 80NSSC18K1729
(Fermi) and NNX17AK42G (ATP), Chandra grant TM9-20002X, and NSF grant
AST-1907955. RP acknowledges support by NSF award AST-1616157.
%

\bibliographystyle{aasjournal}
\bibliography{biblio}

\end{document}